\documentclass[aps,prd,preprintnumbers,amsmath,amssymb,latexsym,nofootinbib,array,enumerate,letter,twocolumn,superscriptaddress]{revtex4-1}

\usepackage{cancel}
\usepackage{here}
\usepackage{comment,braket}
\usepackage{epsf}
\usepackage{amsmath}
\usepackage{graphics}
\usepackage{amsfonts}
\usepackage{amssymb}
\usepackage{latexsym}
\usepackage{color}
\usepackage{natbib}
\usepackage[normalem]{ulem}
\input{colordvi.tex}
\usepackage{feynmp}
\usepackage[utf8]{inputenc}
\DeclareUnicodeCharacter{2212}{-}

\usepackage[pdftex]{graphicx}
\usepackage{bm}
\usepackage[pdftex]{hyperref}
\usepackage[svgnames]{xcolor}
\definecolor{phthaloblue}{rgb}{0.0, 0.06, 0.54}
\hypersetup{
    colorlinks=true,
    linkcolor=phthaloblue,
    citecolor=blue,
    filecolor=blue,
    urlcolor=phthaloblue,
    }

\newcommand{\mev}{\text{MeV}}
\newcommand{\gev}{\text{GeV}}

\renewcommand{\eqref}[1]{Eq.~(\ref{eq:#1})}

\newcommand{\figref}[1]{Fig.~\ref{fig:#1}}

\newcommand{\beq}{\begin{equation}}
\newcommand{\eeq}{\end{equation}}
\newcommand{\beqa}{\begin{eqnarray}}
\newcommand{\eeqa}{\end{eqnarray}}

\begin{document}

\title{
  Dark Photons and Displaced Vertices at the MUonE Experiment
}

\author{Iftah Galon}
\affiliation{New High Energy Theory Center, Rutgers University \\
  Piscataway, New Jersey 08854-8019, USA}
\affiliation{
  Physics Department, Technion—Israel Institute of Technology, \\
  Technion city, Haifa 3200003, Israel
}

\author{David Shih}
\affiliation{New High Energy Theory Center, Rutgers University \\
  Piscataway, New Jersey 08854-8019, USA}

\author{Isaac R. Wang}
\affiliation{New High Energy Theory Center, Rutgers University \\
  Piscataway, New Jersey 08854-8019, USA}

\begin{abstract}
  MUonE is a proposed experiment designed to measure the hadronic vacuum polarization contribution to muon $g-2$ through elastic $\mu-e$ scattering. As such it employs an extremely high-resolution tracking apparatus. We point out that this makes MUonE also a very promising experiment to search for displaced vertices from light, weakly-interacting new particles. We demonstrate its potential by showing how it has excellent sensitivity to dark photons in the mass range $10~\mev \le m_{A'} \le 100~\mev$ and kinetic mixing parameter $10^{-5} \le \epsilon e \le 10^{-3}$, through the process $\mu^{\pm}\, e^- \to \mu^{\pm}\, e^-\, A'$ followed by $A'\to e^+e^-$.
\end{abstract}

\maketitle

\section{Introduction}
\label{sec:intro}
Dark matter is an enduring mystery. Despite ironclad evidence for the gravitational effects of dark matter in astrophysical and cosmological observations, the particle characteristics of dark matter are still obscure.
In fact, dark matter may be part of an intricate dark sector, and the dark sector may communicate with the Standard Model (SM) through ``mediator'' particles.
The masses and couplings of the dark matter and mediator particles are not constrained by ab-initio principles, and can, in theory, span many orders of magnitude. (For comprehensive reviews of dark matter and many original references, see e.g.~\cite{Bertone:2004pz,Freese:2008cz,Hooper:2009zm,Feng:2010gw,Garrett:2010hd,Bauer:2017qwy,Buckley:2017ijx}.)

In recent years, dark matter (and its mediators) in the \mev-\gev\ mass range has received substantial attention~\cite{Essig:2013lka,Alexander:2016aln, Battaglieri:2017aum, Beacham:2019nyx, Alimena:2019zri,Alimena:2021mdu,Bernal:2017kxu,Curtin:2018mvb} as a compelling alternative to more conventional (and increasingly heavily constrained)  dark matter at the weak scale.
Dark sector models, with light mediators of various spin-parity assignments and diverse couplings, give rise to distinctive signatures that are potentially detectable in a variety of experiments, spanning a large range of energies.   Of particular interest are cases in which the mediators are stable, or characteristically long-lived, and propagate macroscopic scales uninterrupted. These can give rise to novel experimental challenges and signatures, e.g.\ displaced decays~\cite{Konaka:1986cb,Riordan:1987aw,Bjorken:1988as,Bross:1989mp,Davier:1989wz,NOMAD:2001eyx,NA64:2018lsq,LHCb:2017trq,Bernardi:1985ny,Blumlein:1990ay,Blumlein:1991xh,CHARM:1985anb,Bjorken:2009mm,Andreas:2012mt,Blumlein:2011mv,Blumlein:2013cua,Gninenko:2012eq,Gninenko:2011uv,CHARM:1985nku,LSND:1997vqj,Essig:2010gu,Williams:2011qb} and missing energy~\cite{BaBar:2017tiz,NA64:2017vtt,DELPHI:2003dlq,DELPHI:2008uka,Essig:2013vha,Davoudiasl:2014kua,NA64:2016oww,Fayet:2007ua}.

In this paper, we explore the sensitivity of the proposed MUonE experiment \cite{Abbiendi:2016xup,Abbiendi:2677471} to long-lived mediators that decay visibly and manifest as displaced vertex signatures.
To be concrete, we demonstrate a proof of concept using the popular ``vanilla'' dark photon model~\cite{Okun:1982xi,Galison:1983pa, Holdom:1985ag, Boehm:2003hm, Pospelov:2008zw}, with displaced decays of the dark photon to $e^+ e^-$.
We show that MUonE tracking capabilities enable a very low-background search for displaced $e^+e^-$ vertices of decay length $\sim 10-100$~mm. In turn, this allows MUonE to have an excellent discovery potential to a wide swath of currently uncovered dark photon parameter space, in the range $m_A\sim 10-100$~MeV and $\epsilon e \sim 10^{-5}-10^{-3}$. (See~\cite{Dev:2020drf,Masiero:2020vxk,Asai:2021wzx} for other studies of the potential of MUonE to be sensitive to new physics, and~\cite{Chen:2017awl,Kahn:2018cqs,Chen:2018vkr,Gninenko:2019qiv,Galon:2019owl} for other proposals to search for light mediators using muon beams.)

\section{The $\text{MUonE}$ experiment}
\label{sec:exp}
MUonE aims to shed light on the discrepancy between the measurement of the muon anomalous magnetic moment, $(g-2)_\mu$~\cite{Muong-2:2006rrc,Muong-2:2021ojo} and the theoretic prediction~\cite{Aoyama:2020ynm}, a $4.2\sigma$ anomaly, starting from an extremely precise measurement of the differential cross-section for the elastic scattering process
\beq
\label{eq:2TO2_process}
\mu^\pm\, e^- ~\to~ \mu^\pm\, e^- \, .
\eeq
This measurement can be used to derive the scale dependence (running) of the electromagnetic fine-structure constant $\alpha$ in the space-like region, which is subsequently used as input for the evaluation of the HVP contribution~\cite{CarloniCalame:2015obs}.  Its concept is based on the NA7 experiment \cite{Amendolia:1984nz,NA7:1986vav}, which measured elastic pion-electron scattering in the 80s. For an up-to-date overview of the technique, the current status of MUonE, and many more original references, see \cite{Abbiendi:2022oks,Pilato:2022dms}.

To carry out this measurement, the MUonE experiment proposes to collide $160$~GeV muons from the CERN M2 beamline with the atomic electrons of thin beryllium targets organized into a series of $40$~consecutive, identical, and aligned modules. Each module consists of one target (15~mm thick), followed by three  $10~\mathrm{cm} \times 10~\mathrm{cm}$ tracking layers. The first tracking layer is located 150~mm after each target. The targets are spaced 100~cm apart.

The planned fixed-target luminosity of the MUonE experiment (taking into account all 40 targets) is~\cite{Abbiendi:2016xup,Abbiendi:2677471}
\begin{align}\label{eq:muonelum}
  {\cal L} = 1.5 \times 10^{4} ~\mathrm{pb}^{-1} \,.
\end{align}
It is projected to take around 3 years for the MUonE experiment to reach this integrated luminosity. Currently, test runs are underway, and an initial run with the full experimental setup is expected to take place before the planned ``Long Shutdown" of the LHC starting in 2026 \cite{Abbiendi:2022oks,Pilato:2022dms}.

To achieve the resolutions required for the precision measurement of the running of $\alpha$, MUonE proposes to use a  CMS-based tracking apparatus~\cite{Abbiendi:2677471,CERN-LHCC-2017-009,Migliore:2797715}, with a resolution on the outgoing angle reaching as low as $0.02~\mathrm{mrad}$.

\begin{figure}[h]
  \centering
  \includegraphics[width=\linewidth]{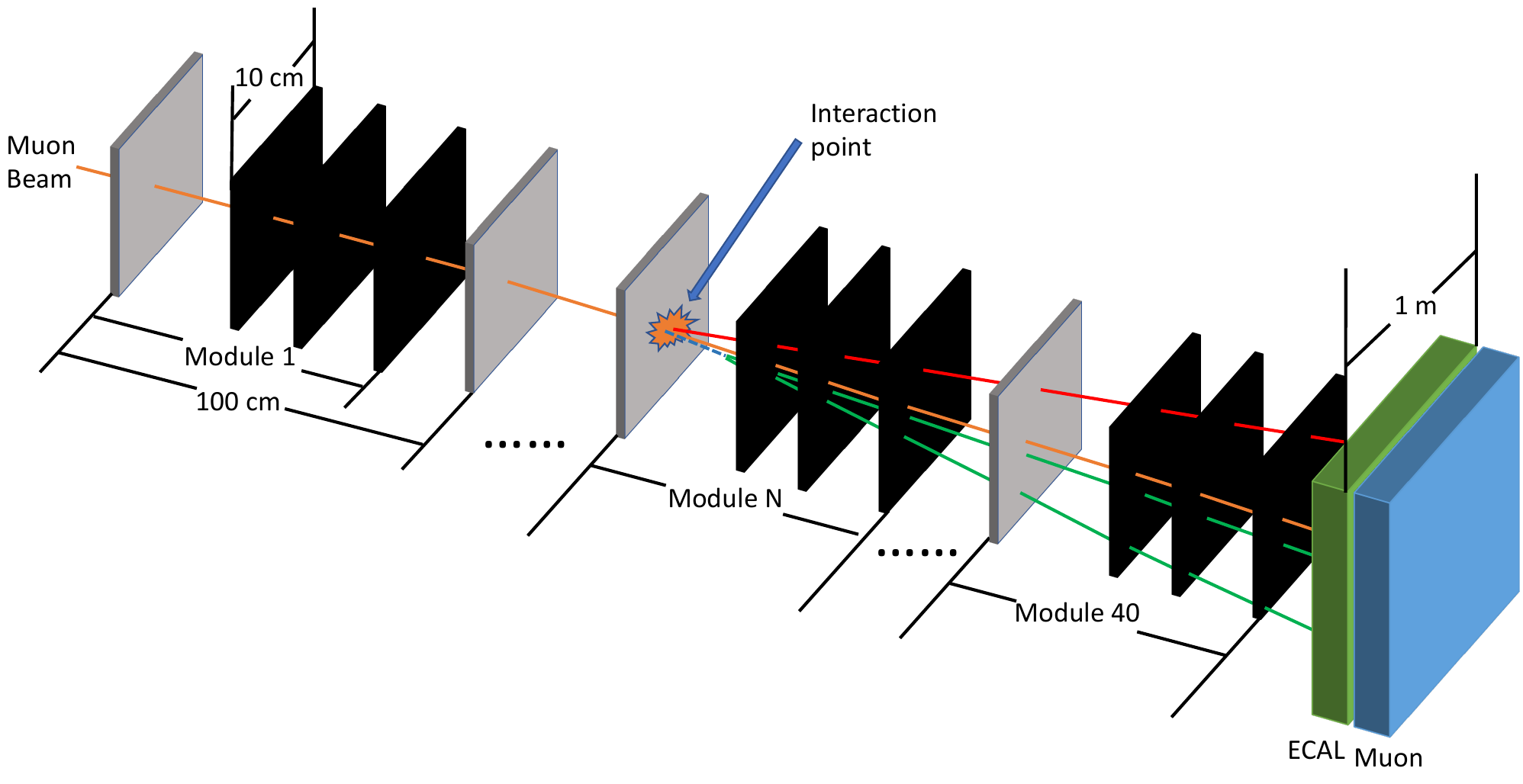}
  \caption{Structure of the MUonE experiment and the 2 to 3 process $\mu^\pm e^- \rightarrow \mu^\pm e^- A^\prime$. Orange lines: incoming muon beam and outgoing muon. Blue dashed line: outgoing dark photon. Red line: outgoing electron from scattering. Green lines: decay products $e^+ e^-$ from the dark photon. Black: tracking layers. Gray: thin $Be$-targets. Green: ECAL. Blue: muon filter. The distance between two adjacent targets is $100~\mathrm{cm}$. Notice that the interaction may happen at any of the 40 targets.}
  \label{fig:experiment}
\end{figure}

Downstream from this setup are located a $1~\mathrm{m} \times 1~\mathrm{m}$ electromagnetic calorimeter (ECAL) and a muon filter, to be used primarily for particle identification (PID).
No magnetic field is employed, so charge discrimination is not possible. Performance numbers, such as energy resolution and PID efficiency, have not been established yet. Below, we will parametrize these and other unknowns in terms of a minimum energy acceptance for electrons and muons, and a maximum allowed PID fake rate for backgrounds. A schematic of the experimental setup is shown in \figref{experiment}.

A displaced vertex signature is defined as the intersection of reconstructed tracks away from the nearest upstream target. Due to its high resolution tracking apparatus, MUonE is extremely efficient in detecting displaced vertices for opening angles between tracks which are as low as $0.1~\mathrm{mrad}$~\cite{Clara:talk}.

\section{The dark photon model}
\label{sec:model}
The ``Vanilla'' dark photon model~\cite{Okun:1982xi,Galison:1983pa, Holdom:1985ag, Boehm:2003hm, Pospelov:2008zw} is a popular simplified model which is used as a benchmark scenario for exploring light, weakly-coupled, long-lived particles~\cite{Essig:2013lka,Alexander:2016aln, Battaglieri:2017aum, Alimena:2019zri}.
In this model, a dark sector Abelian gauge symmetry, $U(1)_D$, is broken, and kinetically mixes with the SM hypercharge gauge factor.
Dark gauge bosons that are much lighter than the weak scale, predominantly mix with the
unbroken $U(1)_{EM}$ factors, and their induced couplings are photon-like~\cite{Curtin:2014cca, Feng:2016ijc, Izaguirre:2017bqb}.
Focusing on this, and denoting the dark photon, and its field strength by $A'$, and $F'$, respectively, the model reads
\beq
{\cal L}_{A'} = -\frac14 F'_{\mu\nu}F'^{\mu\nu} -\frac12 m_{A'}^2  A'_\mu  A'^\mu
-\epsilon e A'_\mu J_{EM}^\mu
\eeq
where the dark photon interaction with the electromagnetic current, $J_{EM}$ is suppressed by the kinetic mixing parameter, $\epsilon$, $J_{EM}^\mu = \sum_k Q_k \bar\psi_k \gamma^\mu \psi_k$, and $Q_k$ is the electric charge of the fermion field $\psi_k$ in units of $e$.

In this work, we assume that the dark photon decays predominantly visibly to $e^+e^-$ pairs.
Such a scenario arises naturally for a dark photon mass in the range $2m_e \le m_{A'} < 2 m_\mu$, when the dark photon decays to the dark sector are suppressed (due to a small coupling or phase-space). The decay width of the dark photon to $e^+e^-$ is given by
\beq
\label{eq:Gamma_Ap}
\Gamma_{A'} = \Gamma(A' \to e^+e^-)= \frac{(\epsilon e)^2}{12\pi}m_{A'}\left(
1 + 2 \frac{m_e^2}{m_{A'}^2}
\right)
\sqrt{
  1 - 4\frac{m_e^2}{m_{A'}^2}
}
\eeq
The dark photon decay length (in the lab frame) is given by
\begin{align}
  \bar d_{A'}
   & =\frac{\gamma_{A'}\beta_{A'}}{\Gamma_{A'}} = \frac{p_{A'}}{m_{A'} \Gamma_{A'}}\nonumber \\
   & \approx
  32~\mathrm{mm}
  \left(\frac{10^{-4}}{\epsilon}\right)^2
  \left(\frac{p_{A'} }{ 10~\gev}\right)
  \left(\frac{ 50~\mev}{m_{A'}}
  \right)^2
  \label{eq:dbarAp}
\end{align}
where we use the characteristic dark photon momentum, $p_{A'} \simeq 10~\gev$, see Appendix~\ref{sec:PSconstraints} for details. This means the dark photon typically decays between the target and the first tracking layer at MUonE.

In the decay $A' \to e^+e^-$, given that $A'$ is sufficiently energetic, the characteristic opening angle between the electron and positron is given by
\beq
\label{eq:opening angle}
\theta_{ee} \simeq \frac{2 m_{A^\prime}}{E_{A^\prime}}
\eeq
For dark photons with mass $\sim 50~\mathrm{MeV}$ and typical energy $10~\gev$, the opening angle is around $10~\mathrm{mrad}$. Given the upper bound on the dark photon angle (which also happens to be ${\mathcal O}(10)~\mathrm{mrad}$ for these dark photon masses, see Appendix~\ref{sec:PSconstraints}), this makes the decay products very likely to pass through the tracking layers and the ECAL/muon system.

All in all, combining \eqref{dbarAp} and \eqref{opening angle}, we conclude that the geometry and design of MUonE are well-suited to search for dark photons with displaced decays to $e^+e^-$ pairs. In the next section, we will propose a number of search strategies that leverage MUonE capabilities to identify displaced vertices, determine particle species, and reject various potential backgrounds.

\begin{figure}[t]
  \centering
  \includegraphics[width=\linewidth]{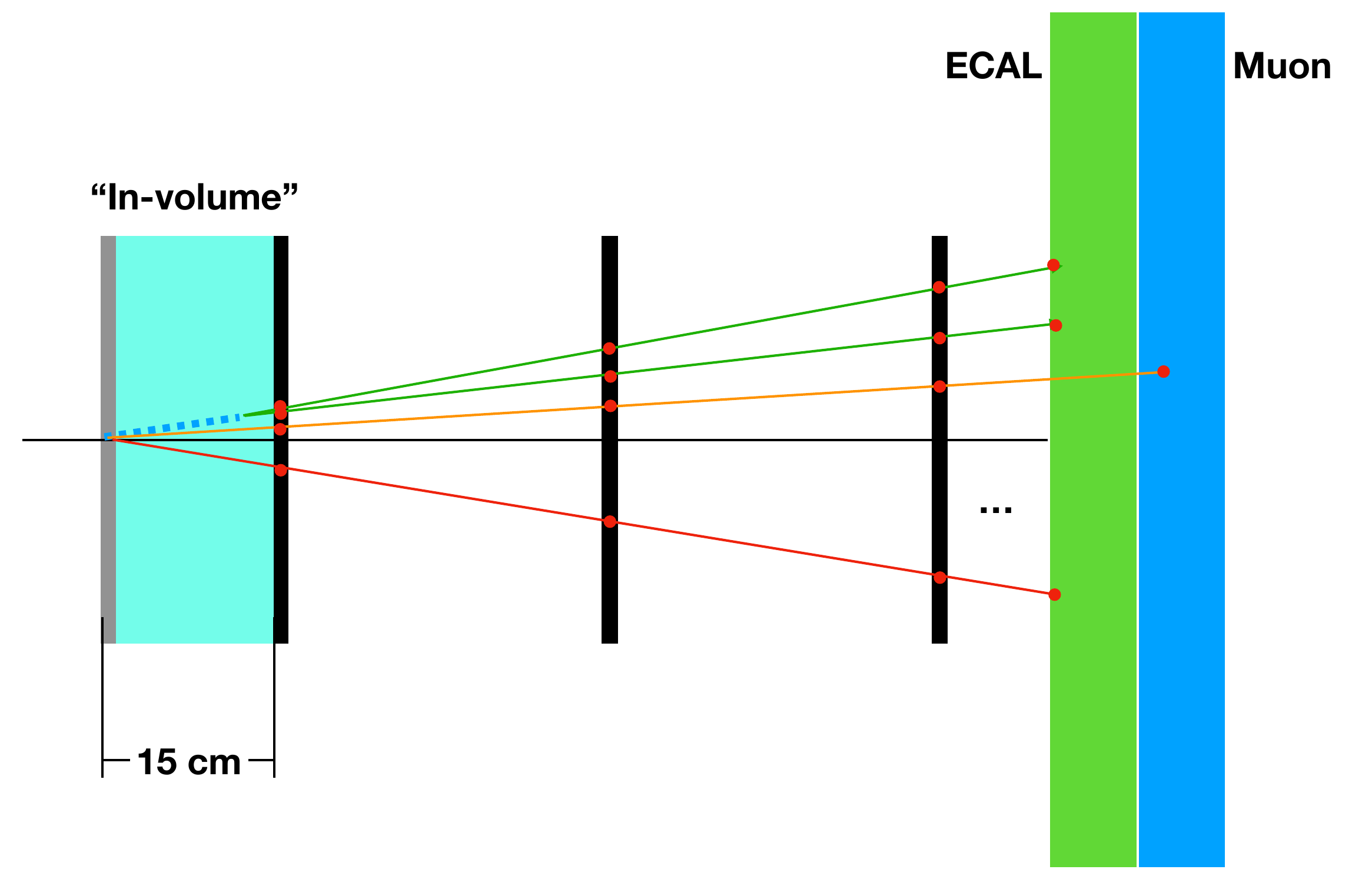}
  \caption{Close look at the decay volume and angular acceptance. Gray: target where the collision happens. Black: tracking layers. Blue dashed line: dark photon trajectory. Green lines: decay products $e^+ e^-$ pair. Orange line: outgoing muon. Red line: Outgoing electron from scattering at the target. Cyan region: 'in-volume' requirement region. Red dots: hits left by charged leptons in the detectors.}
  \label{fig:close}
\end{figure}

\section{The displaced vertex search}
\label{sec:search}

\subsection{Search strategies}
\label{sec:strategies}
We propose to use MUonE to search for dark photons that are produced in $\mu\, e$ scattering:
\beq
\label{eq:2TO3_process}
\mu^\pm\, e^- ~\to~ \mu^\pm\, e^- A'
\eeq
and subsequently decay to $e^+e^-$ pairs.

We require events to contain a pair of charged tracks consistent with production from a displaced $A'\to e^+e^-$. To enable accurate tracking and particle ID, we require these tracks to pass through the three tracking layers in a module, and make it all the way to the ECAL. As discussed in previous sections, this requirement has an ${\mathcal O}(1)$ acceptance at MUonE.  To further ensure that the displaced $e^+e^-$ pair is accurately reconstructed, we apply a cut on their lab frame opening angle of $\theta_{ee} \ge 1~\rm mrad$.

The reconstructed tracks must point towards a common vertex that is significantly displaced from any potential scattering material (the target or the tracking layers). To guarantee this, given an expected resolution along the $z$-direction of the displaced vertex location of $\delta z = 1~\mathrm{mm}$~\cite{Clara:talk}, we require a reconstructed vertex that is at least $10\delta z$ away from the target and from the first tracking layer. Following the geometry shown in Fig~\ref{fig:close}, we take the fiducial volume for the dark photon decay to be
\beq
L_{\rm min} = 25~\mathrm{mm} \,,
\qquad
L_{\rm max} = 140~\mathrm{mm}
\eeq
as measured from the beginning of each module.

In principle, beam muons can scatter off the target atomic electrons in any one of the 40 MUonE modules, but energy loss when traversing multiple targets is a potential issue. For example, an electron with energy 100 MeV, 1 GeV and 5 GeV will lose approximately 8\%, 5\% and 4\% of its energy when passing through a single beryllium target.\footnote{Energy loss of the primary muon is much less of an issue. The muons always have energy larger than $10~\gev$, see Appendix~\ref{sec:PSconstraints}. A muon with an energy $10~\gev$ will lose approximately 0.05\% of its energy when passing through a single target, and muons with higher energies will lose less~\cite{Workman:2022ynf,groom2001muon}.} For details, see~\cite{Workman:2022ynf,walters2017stopping}. The accuracy of the tracking system and particle ID in the ECAL/muon filter may be degraded due to this energy loss, and the lowest energy for the particle to be identified by the ECAL also remains unknown. To minimize this energy loss, only the last 5 modules are used in this proposed search.

Furthermore, to model the reduced detector acceptance due to energy loss, we impose a simple minimum energy threshold on all the particles in the event. Due to the uncertainty of the detector behavior, we consider 3 different minimum energy thresholds, -- $100~\mathrm{MeV}$, $1~\gev$ and $5~\gev$ -- in our study. We stress that this is not an analysis-level cut -- we are not assuming MUonE will be able to accurately measure these energies. Rather, we are using these minimum thresholds as a proxy for the detector acceptance, i.e.\ we assume particles with energy less than these thresholds will typically be lost or badly reconstructed, leading the event to be rejected.

The primary muon always has a small angle less than 10~mrad (see Appendix~\ref{sec:PSconstraints}), so it is guaranteed to pass through the three tracking layers, and also to reach the ECAL/muon filter.  This makes it realistic to identify the primary muon and measure the angle, and we lose no signal rate in requiring the presence of the primary muon (with an angle less than 10~mrad) in every event. This event topology (two tracks consistent with displaced $e^+e^-$ and one track consistent with the primary muon) will be referred to as the ``3-track search".

To further reduce the background (see the next subsection), one can further require the primary electron to pass through the tracking layers and the ECAL. We call this the ``4-track search''. However, the primary electron is generally the lowest energy particle in the event and can be subject to significant energy losses. Thus the number of signal events might be more sensitive to the detailed modeling of the detector response.

\subsection{Backgrounds}
\label{sec:backgrounds}
A comprehensive estimation of the rates for background processes at MUonE requires dedicated simulation tools (for example a Geant4~\cite{agostinelli2003geant4,allison2006geant4,allison2016recent} model of MUonE), and is therefore well beyond the scope of this work.
Nevertheless, it is instructive, as a preliminary estimate, to enumerate and discuss the potential backgrounds of the displaced vertex search. These backgrounds typically fall into two primary categories:
(a) SM processes with an inherently displaced vertex, and (b) SM process in which a prompt vertex fakes a displaced vertex due to tracker resolution effects.

In the first category, the displaced vertex signature arises from the production of long-lived SM particles such as neutral Kaons, which can travel and decay to a  $\pi^+ \pi^-$ pair that could potentially fake the electron/positron pair.
At MUonE, such particles would be produced in muon-nucleon scattering. To simulate these background processes,
we use the \verb|softQCD| functionality of  \textsc{pythia8.307}~\cite{Bierlich:2022pfr} to simulate muon-proton collision.
This includes all the soft QCD processes in a full range of $Q^2$ that can be a good complement to the deep inelastic scattering (large $Q^2$ only)\footnote{The DIS processes can be easily rejected as they always lead to a primary muon with angle larger than 10~mrad, whereas primary muons from the dark photon signal always have angle less than 10~mrad. For this reason, we only consider other processes.}.
As described in the main text, we require events to have exactly 3 or 4 charged tracks, with one or two originating from the target and two originating from a displaced decay in the fiducial volume. We reject any events that have additional tracks that hit even a single tracking layer, or neutral particles that hit the ECAL.
To achieve a comprehensive picture of the possible background events (and a reasonably accurate estimate of their numbers), we simulated 15B  \verb|softQCD| events, which is about 10 times more events of this type than expected from the experiment.

Based on our simulation, we expect approximately 4,000 events can fake the signal topology of the ``3-track search'' before particle ID, in the planned lifetime of the MUonE experiment. The vast majority of these
arise from displaced $K^0 \rightarrow \pi^+ \pi^-$ and $\Lambda^0 \rightarrow p \pi^-$, and so have two displaced tracks that must fake an electron/positron pair. These events can be rejected by the ECAL/muon PID system, as long as the per-particle-fake rate is less than 1.6\%.
A much smaller number (just 8 events expected) have one real electron/positron and one other particle (usually a charged pion) faking a displaced track; these can arise from rare decays such as  $K^0 \rightarrow  \pi^\pm e^\mp \nu_e$ and $\Lambda^0\to p e^-\bar\nu_{e}$. Finally, just 0.5 events are expected to contain a genuine displaced $e^+e^-$ pair. These can arise from  decays like $K^0\to \pi^0\pi^0$, with one pion decaying to $e^+e^-\gamma$, and with only the electron-positron pair recorded by the detector. Fortunately, we find that all genuine displaced $e^+e^-$ events have at least one non-electron particle hitting the ECAL (usually a charged pion from the primary interaction). So these few events with one or two real displaced electron/positrons should all be easily rejected  by the ECAL/muon PID system.

Meanwhile, in the ``4-track search" strategy where the presence of the primary electron is also required (note: this is a strict subset of the ``3-track search", which allows for the primary electron but does not require it), we find the number of background events before PID is expected to be around 2,000, the vast majority of which again have no $e^\pm$ in the final state. Therefore, they can all be safely rejected as long as the per-particle-fake rate is less than 8.0\%.

\begin{figure*}[t]
  \centering
  \includegraphics[width=0.48\linewidth]{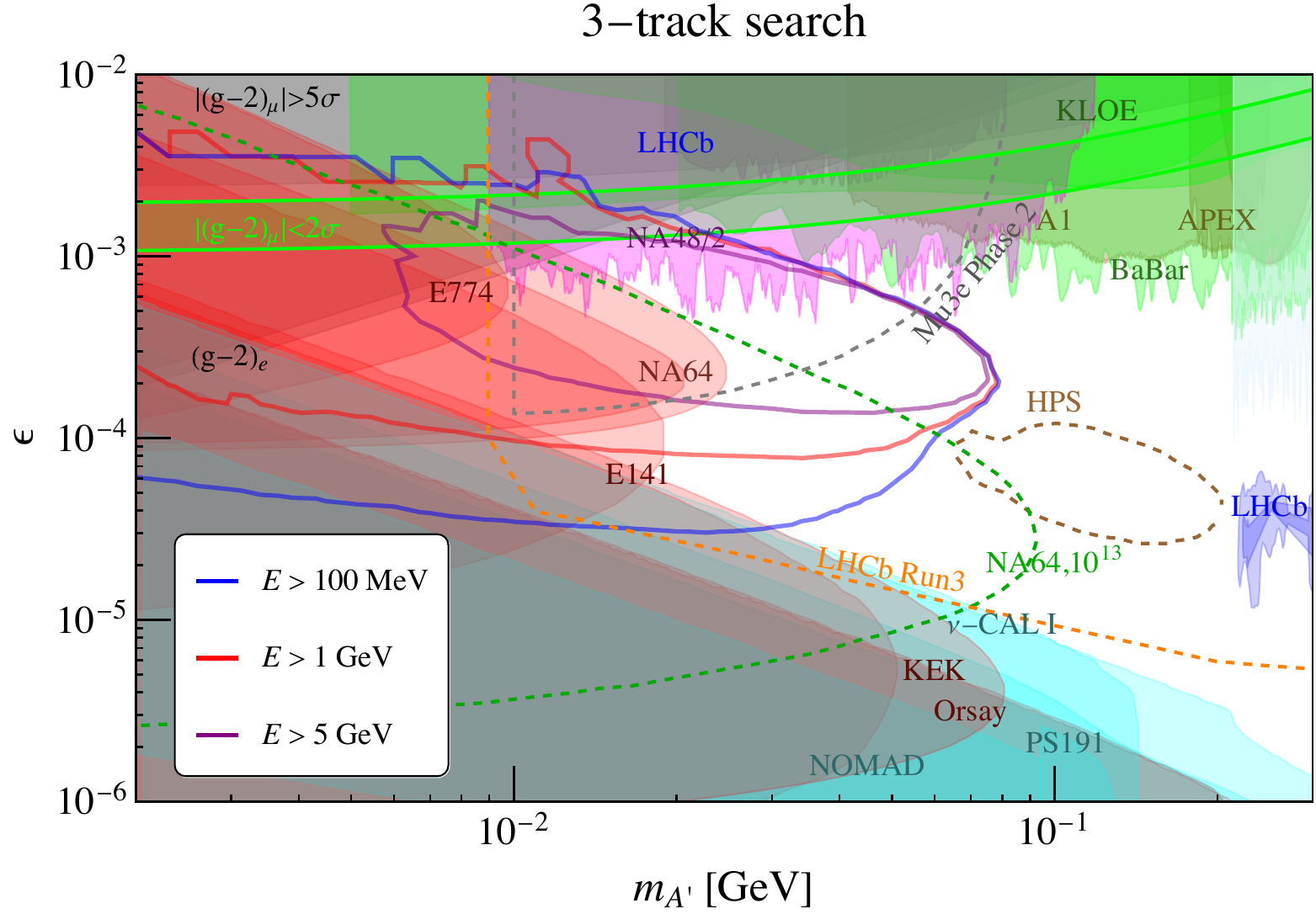}
  \hspace{1em}
  \includegraphics[width=0.48\linewidth]{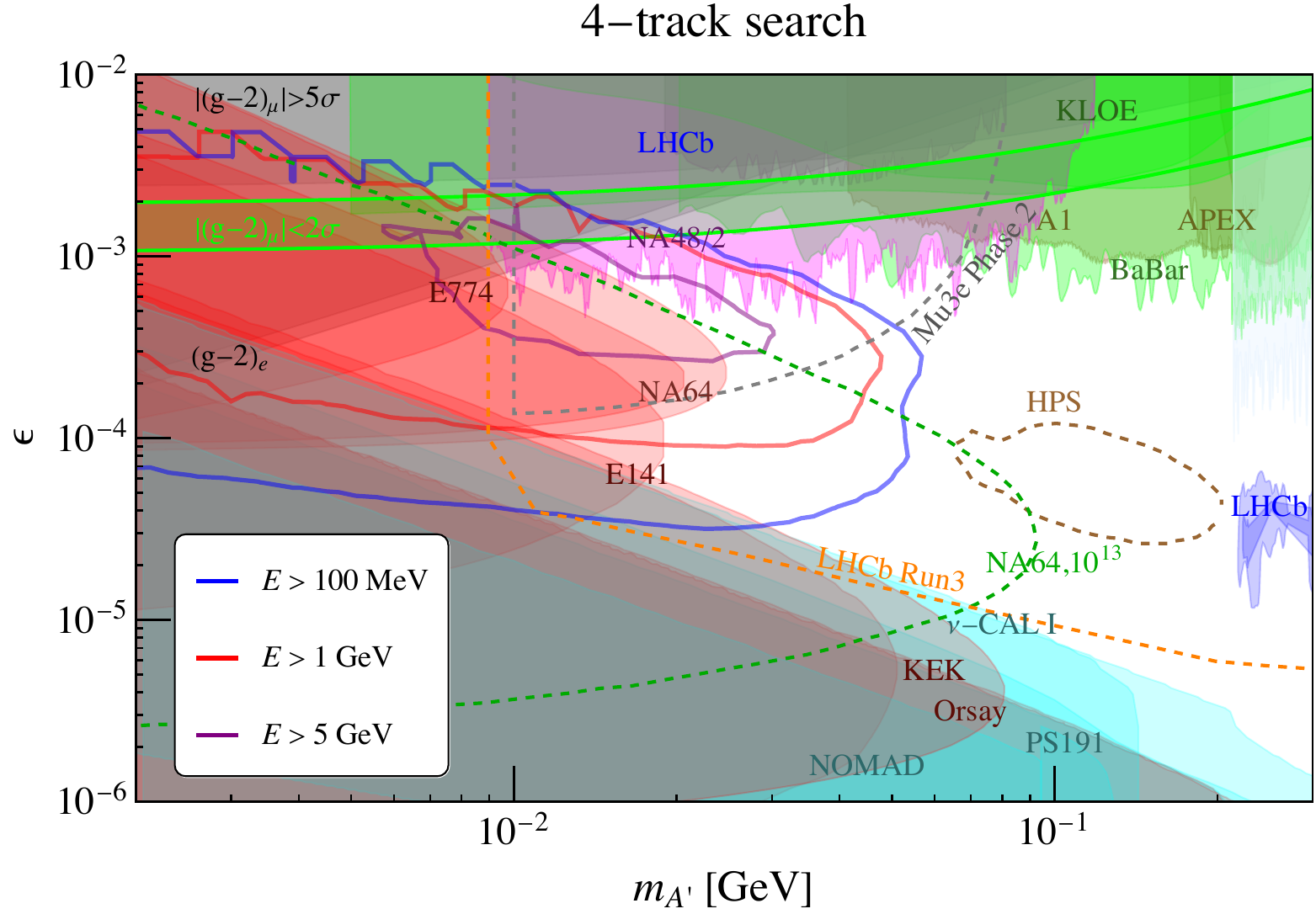}
  \caption{Exclusion curves at $95\%$ C.L.(red). Shaded regions are excluded by other experiments. MUonE results in different strategies in this work are shown by solid lines. Future projections from HPS~\cite{HPS:2018xkw,Baltzell:2022rpd}, the LHCb upgrade~\cite{Ilten:2015hya,Ilten:2016tkc}, Mu3e~\cite{Echenard:2014lma} and NA64e++ \cite{Gninenko:2300189} are shown by dashed lines. Left panel: 3-track search. Right panel: 4-track search. Blue: energy threshold 100 MeV. Red: energy threshold 1 GeV. Purple: energy threshold 5 GeV.
  }
  \label{fig:reach plot}
\end{figure*}

In the second category, the displaced vertex signature arises from SM processes
which are prompt, i.e. the production of the background event occurs at the target (for a recent calculation of background process at NNLO relevant to this proposed search, see~\cite{Budassi:2021twh}). Due to tracker resolution effects, tracks in the event may reconstruct to a vertex that manifests as displaced. This category includes processes which share the same particle content as the signal, such as $\mu\,e\to \mu\,e\gamma^*\,(\gamma^*\to e^+e^-)$, as well as processes such as $\mu\,e\to \mu\,e\, C^+C^-$, where $C^\pm$ is any charged particle that can fake an electron signature. We estimate that the decay-in-volume requirements (recall, we are requiring the displaced vertex be reconstructed at least 10$\delta z$ away from the target or tracking layer boundaries) combined with particle-identification requirements should be more than sufficient to
completely eliminate such backgrounds. A more precise estimate, that takes into account e.g.\ the effect of non-Gaussian tails, would require a full and detailed detector simulation. But even if these types of background events are more than expected from the naive Gaussian resolution, we note that the fiducial volume for displaced decays can be further reduced without suppressing the reach too much -- for example, changing the cut from 10$\delta z$ to 25$\delta z$ will shrink the maximum mass we can probe by only $15~\mathrm{MeV}$.

In the remainder of this analysis, we assume that background rates are negligible.

\section{Analysis and results}
\label{sec:result}
In this work, we simulate dark photon events using \textsc{MadGraph5\_aMC@NLO v3.4.1}~\cite{Mattelaer:2021xdr,Alwall:2011uj,Alwall:2014hca,Frixione:2021zdp}, with UFO file implemented in \textsc{FeynRules 2.3.32}~\cite{Degrande:2011ua,Alloul:2013bka}. For the mass range $[2~\mev,200~\mev]$ that we are interested in, the cross section $(\epsilon e)^{-2} \sigma$ varies from $10^9~\rm pb$ to $10^5~\rm pb$, decreasing as mass increases. (See also~\cite{Ilten:2016tkc} for a data-driven approach to dark-photon production cross section estimation.)
Samples of 100K $\mu e \rightarrow \mu e A^{\prime}$ events are generated for each dark photon mass point. Dark photon decays to $e^+e^-$ are simulated by hand using Eq.~(\ref{eq:Gamma_Ap}).
Event selection is performed for the search strategies described in Section~\ref{sec:strategies}.
The dark photon event yield following these cuts is given by
\begin{align}
  \label{eq:numbers}
  N =  {5\over 40}\mathcal{L} \cdot \sigma \cdot \epsilon_{\rm signal}\,,
\end{align}
where $\epsilon_{\rm signal}$ is the fraction of events that satisfy our requirements, the integrated luminosity is given in Eq.~(\ref{eq:muonelum}), and the factor $5/40$ comes from the fact we are only using the last 5 modules.

For each mass point, the dark photon production cross section, $\sigma$, is estimated in simulation with $\epsilon e = 1$, and is rescaled by $e^2 \epsilon^2$ for each $(\epsilon,\, m)$ parameter-space point. Similarly, the decay in volume probability, is estimated for each parameter-space point using the dark photon decay-length, \eqref{dbarAp}.

Assuming zero background events, the MUonE sensitivity reach at $95\%$ C.L. is shown in~\figref{reach plot} for the different search strategies and energy thresholds.
Signal event contours can be found in Appendix~\ref{sec:contours}. The existing limits (from~\cite{Merkel:2014avp,Bodas:2021fsy, Abrahamyan:2011gv,BaBar:2014zli, TheBABAR:2016rlg, Ablikim:2017aab,CHARM:1985anb,Tsai:2019mtm,CMS:2019kiy,Andreas:2012mt,Bjorken:2009mm,Riordan:1987aw,Bross:1989mp,Adrian:2018scb,Konaka:1986cb,Anastasi:2015qla,Anastasi:2016ktq,Anastasi:2018azp,Babusci:2012cr,Babusci:2014sta,LHCb:2017trq,Aaij:2019bvg,NA482:2015wmo,NA64:2018lsq,Banerjee:2019hmi,NOMAD:2001eyx,Blumlein:1990ay,Blumlein:1991xh,Tsai:2019mtm,Davier:1989wz,Bernardi:1985ny}) are reconstructed via \textsc{darkcast}~\cite{Ilten:2018crw}.

There are several other experimental proposals to probe the dark photon in this parameter space, e.g. HPS~\cite{HPS:2018xkw,Baltzell:2022rpd}, the LHCb upgrade~\cite{Ilten:2015hya,Ilten:2016tkc}, Mu3e~\cite{Echenard:2014lma} and NA64e++ \cite{Gninenko:2300189} (for a review, see \cite{Lanfranchi:2020crw}).
For comparison, we also show these projections in Fig~\ref{fig:reach plot}.
From this figure, we see that the MUonE reach is complementary (covers different parameter space) to that of HPS, Mu3e and NA64e++. Meanwhile it is a subset of the claimed reach of LHCb upgrades. However, we note that LHCb is a very different (messier, hadronic) environment, with a completely different set of backgrounds and systematics, so even if their projections prove to be accurate, having another probe of the same physics with MUonE could be valuable.

\section{Conclusions}
\label{sec:conclusions}
In this work, we have shown that the design of the MUonE experiment makes it a promising environment to search for light mediators with visibly displaced decays.
Focusing on the vanilla dark photon model for a proof-of-concept, we have identified $\mu+e\to \mu+e+A'$ with $A'\to e^+e^-$ as a promising final state, and we have argued that backgrounds are likely to be negligible.  By searching for visibly displaced $e^+e^-$ pairs, possibly in conjunction with $\mu$ and $e$ from the initial $2\to 3$ scattering process, we have demonstrated that MUonE could have excellent sensitivity to a crucial range of dark photon parameter space ($m_{A'}$ between 10-100~MeV and $\epsilon e$ between $10^{-5}-10^{-3}$).

Of course, this being just a proof-of-concept study, much work remains to accurately estimate background and signal yields; this likely requires detailed GEANT4 simulations that are beyond the scope of this work.
Also, it would be interesting to consider other light mediator models that MUonE could be sensitive to besides vanilla dark photons, such as ALPs and leptophilic mediators. Leptophilic models would be an especially attractive target for MUonE, since many of the bounds shown in~\figref{reach plot} rely on couplings of dark photons to quarks and hence would disappear for such models. In any case, we hope this work provides further motivation for the MUonE proposal and illustrates how it could have multiple purposes beyond its original goal of measuring HVP for muon $g-2$.



\begin{acknowledgments}

We are especially grateful to Umberto Marconi, Clara Matteuzzi and Giovanni Abbiendi for their encouragement throughout the course of this project and for crucial discussions about the setup of the MUonE experiment, backgrounds and search design. In addition, we thank Simon Knapen, Paride Paradisi, Yotam Soreq, Robert McGehee, Xun-Jie Xu, Zhite Yu, Felix Kling and Gordan Krnjaic for helpful discussions. Finally, we are extremely grateful to Clara Matteuzzi and Giovanni Abbiendi for detailed feedback on the draft.
This work was supported by DOE grant DE-SC0010008.
The work of IG is supported in part by the Israel Science Foundation (Grant No. 751/19), and by BSF-NSF grant 2020-785.

\end{acknowledgments}

\newpage
\appendix

\section{Phase-Space Constraints in The $2\to3$ Process $\mu^- e^- \to \mu^- e^-  A'$}
\label{sec:PSconstraints}
The phase-space constraints on the angle and energy
of the outgoing particle $X=e, \mu,\, A'$, in the $2\to3$ scattering process  $\mu^- e^- \to \mu^- e^-  A'$ at MUonE are given by
\beqa
\label{eq:angle_energy_constrain}
&&\theta_X^{out} < \arccos\left(\sqrt{\frac{a^2 - b_X^2}{a^2-1}}\right)
\\
&&E_X^- \le E_X^{out}  \le E_X^+
\\
&&E^\pm_X = m_X\left(
a\, b_X \pm \sqrt{(a^2-1)(b_X^2-1)}
\right)
\eeqa
where
\beqa
a &=& \gamma_{CM} = \frac{E_\mu^{in} + m_e}{\sqrt{s_{CM}}}
\\
b_{X=e} &=& \frac{s_{CM} + m_e^2 - (m_\mu + m_{A^\prime})^2}{2m_e\sqrt{s_{CM}}}
\\
b_{X=\mu} &=& \frac{s_{CM} + m_\mu^2 - (m_e + m_{A'} )^2 }{2m_\mu\sqrt{s_{CM}}}
\\
b_{X=A'} &=& \frac{s_{CM} + m_{A'}^2 - (m_e + m_\mu )^2 }{2m_{A'}\sqrt{s_{CM}}}
\eeqa
and we have used
\beq
s_{CM} = m_e^2 + 2m_e E_\mu^{in} +m_\mu^2
\eeq
\begin{figure*}[h]
  \includegraphics[width=0.9\linewidth]{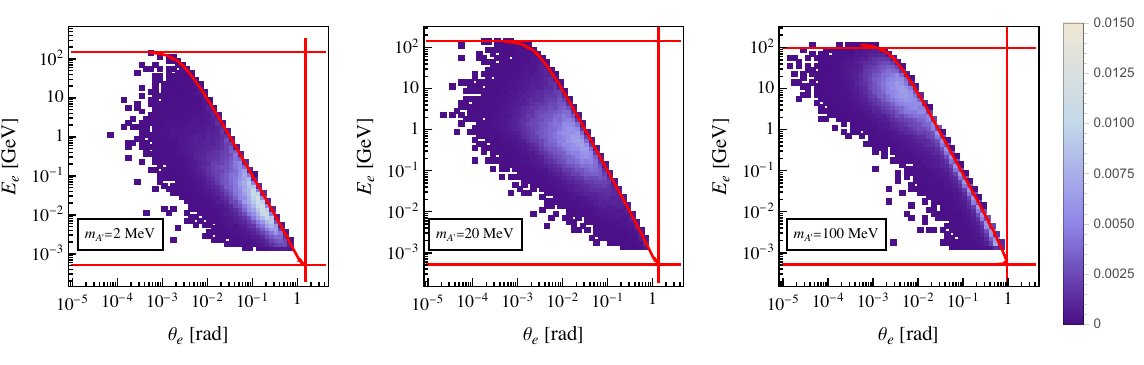}
  \includegraphics[width=0.9\linewidth]{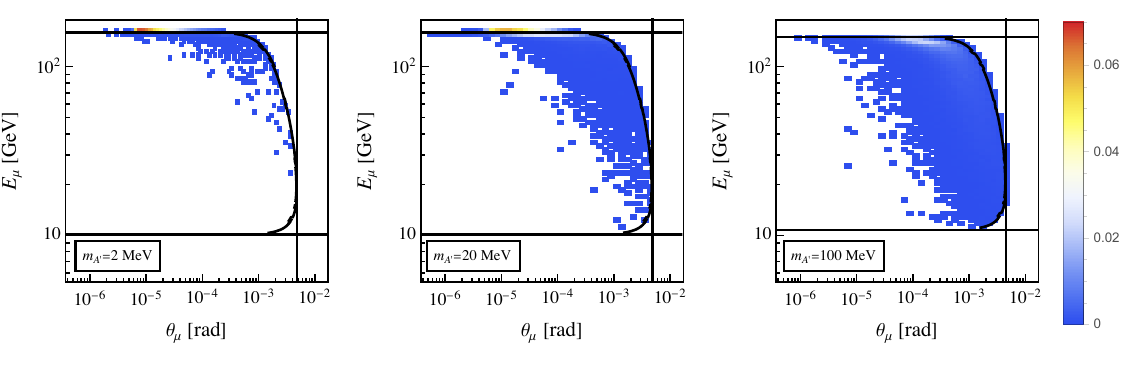}
  \includegraphics[width=0.9\linewidth]{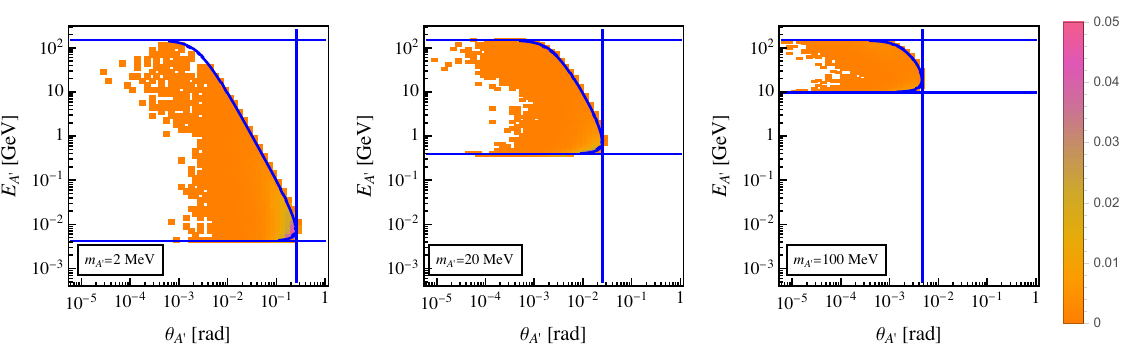}
  \caption{Density plots for the energy and outgoing angle of the electron, muon, and the dark photon for dark photon mass $m_{A^{\prime}} = 2, 20, 100~\mev$ and coupling $\epsilon e = 1.0$. Vertical and horizontal lines: limits on the maximal/minimal angle and energies from the phase space constraints. Curves: phase space constraints.}
  \label{fig:2dhist}
\end{figure*}

We note that these arise at MUonE because the muon projectile is heavier than the target electron~\cite{Galoninprogress}. Density plots in the energy vs.\ angle plane for the outgoing muon, electron, and the dark photon are shown in \figref{2dhist}.

\begin{figure*}[h]
  \centering
  \includegraphics[width=3.5in]{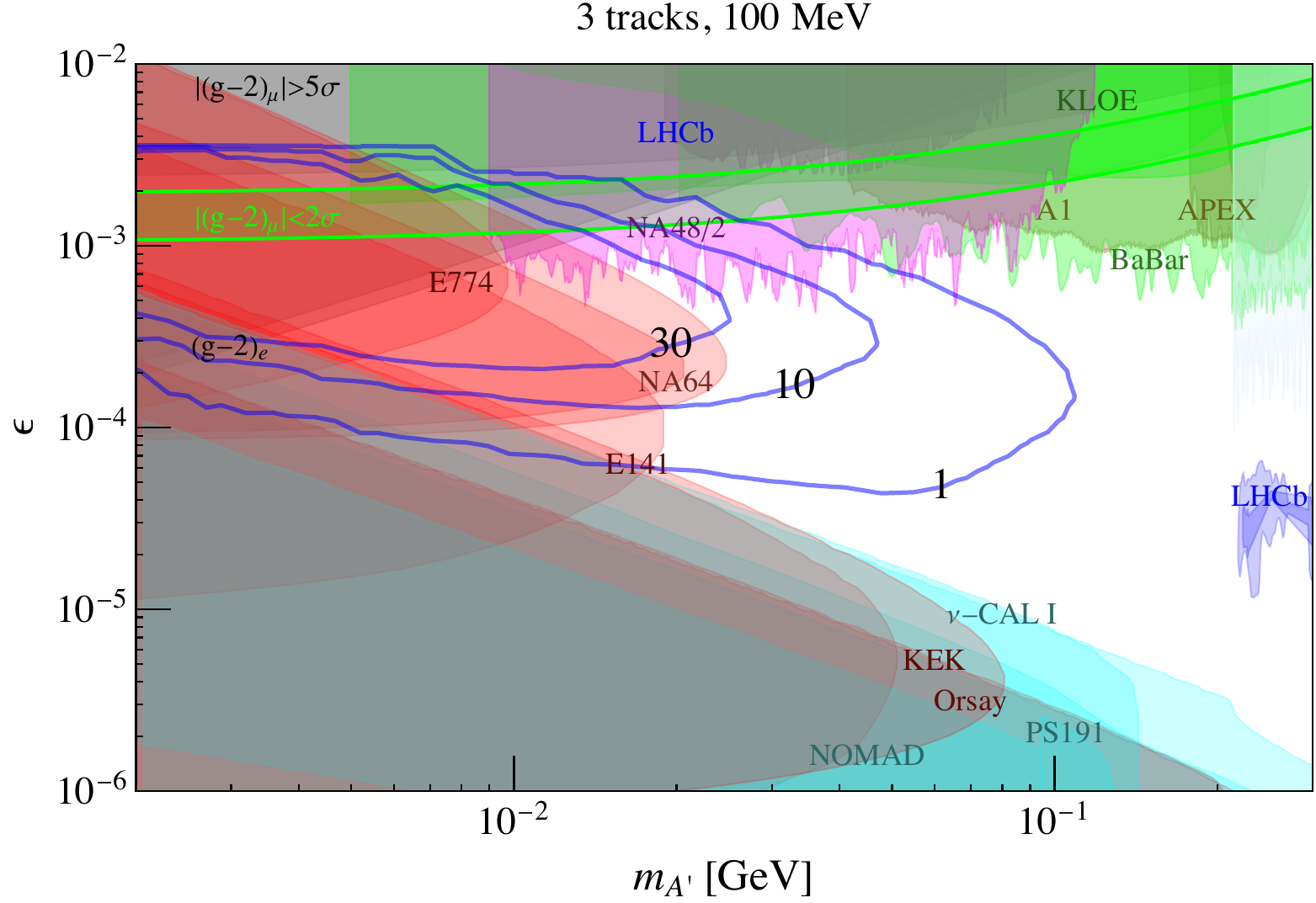}
  \hfill
  \includegraphics[width=3.5in]{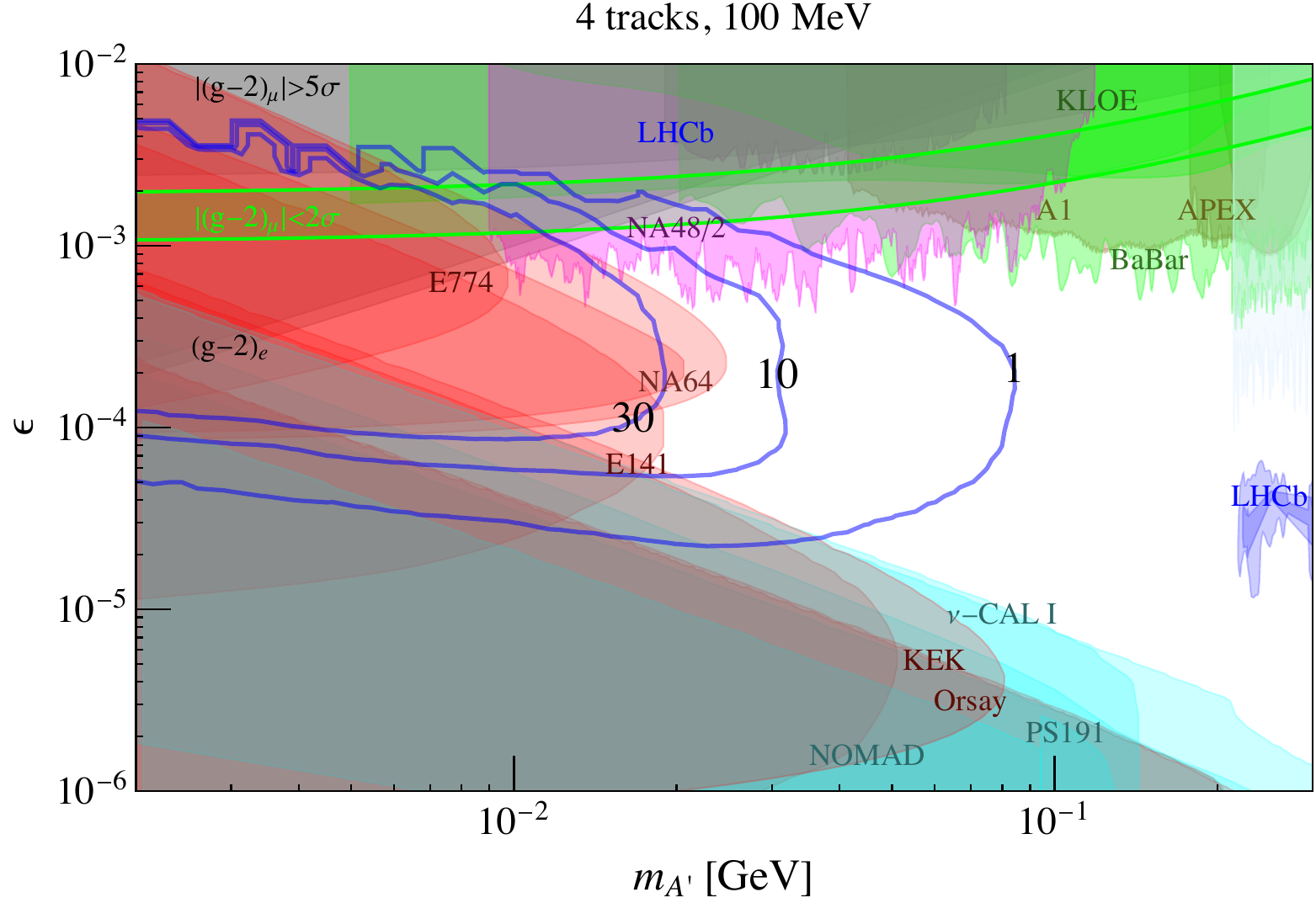}

  \vspace{1em}

  \includegraphics[width=3.5in]{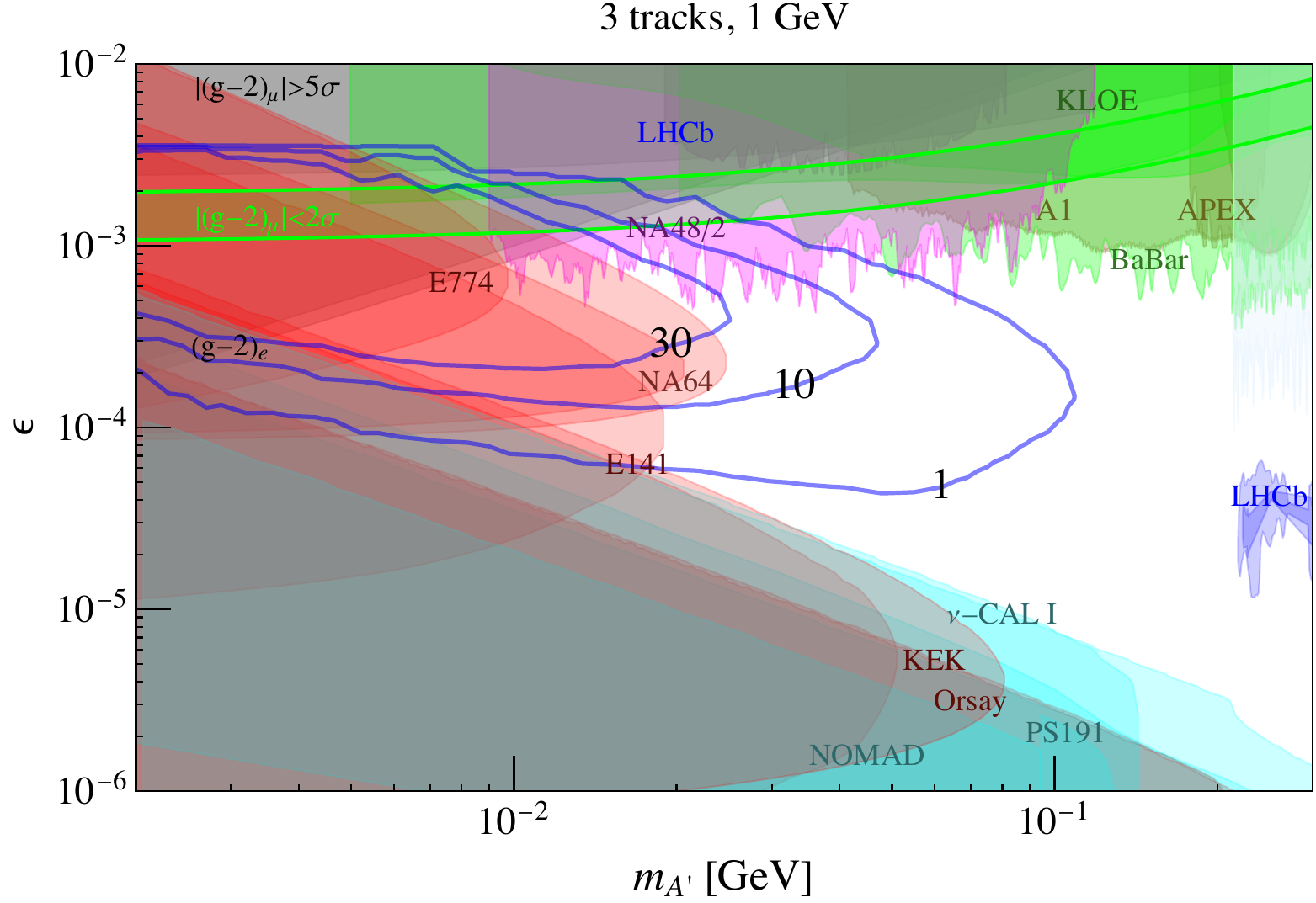}
  \hfill
  \includegraphics[width=3.5in]{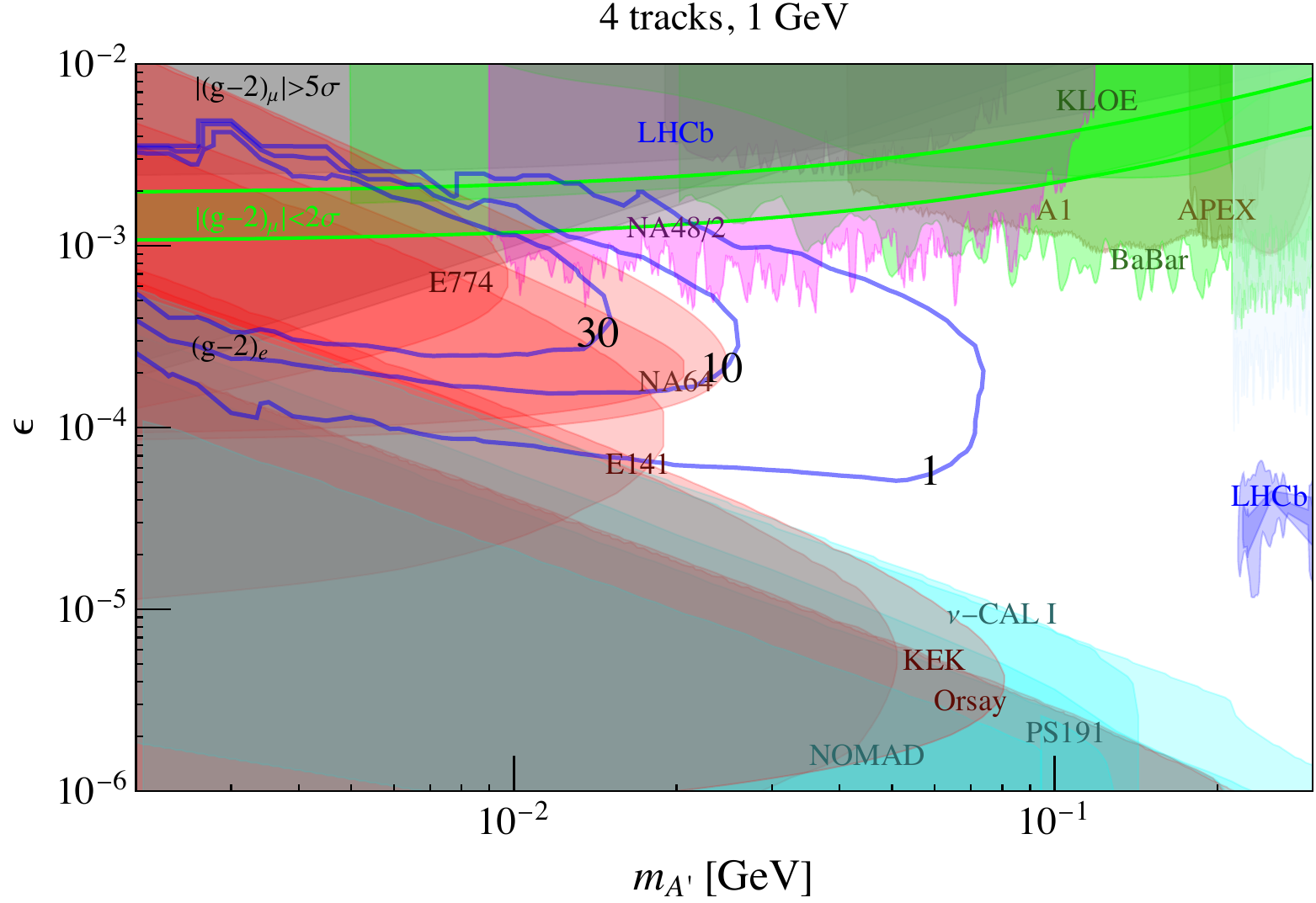}

  \vspace{1em}

  \includegraphics[width=3.5in]{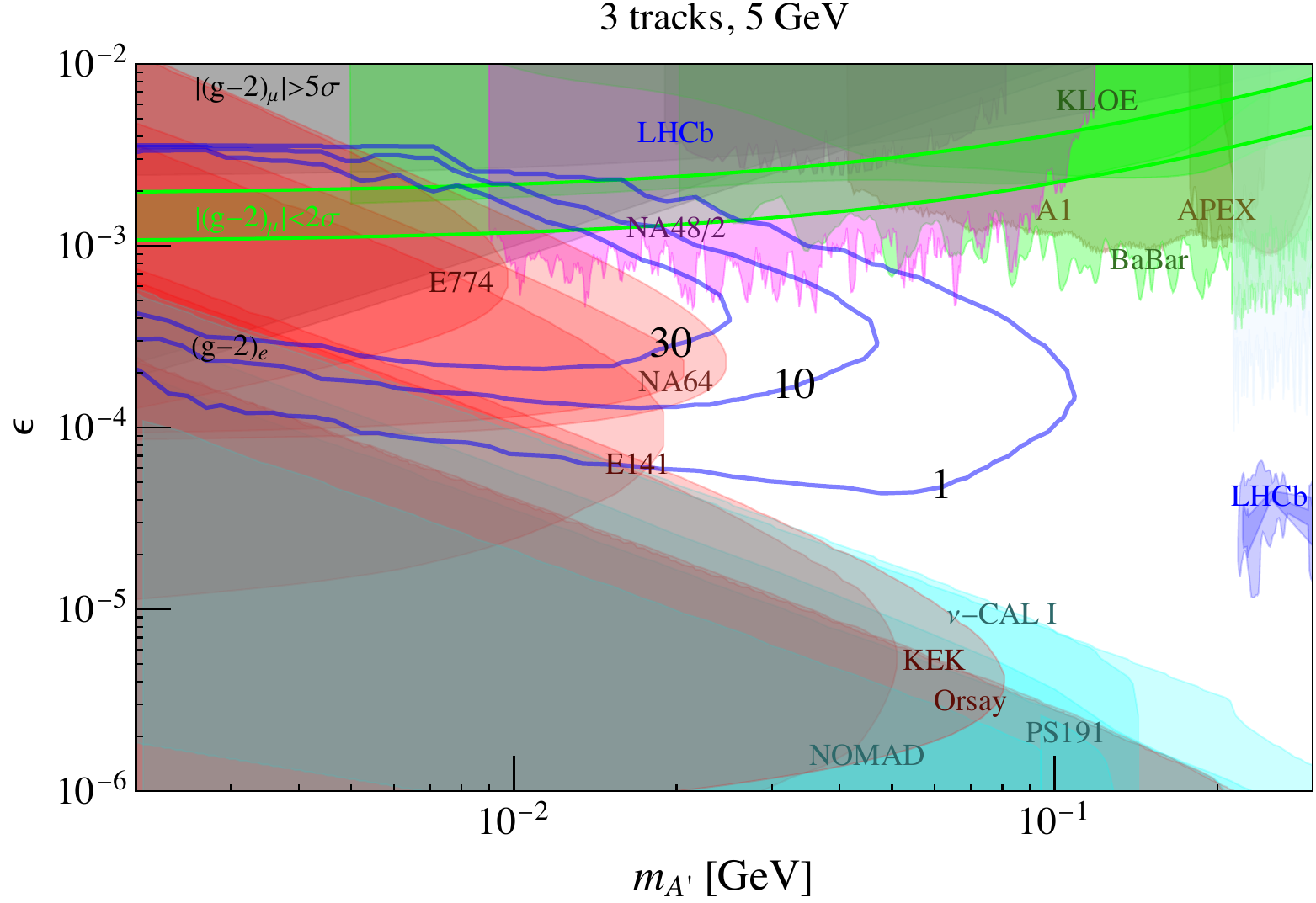}
  \hfill
  \includegraphics[width=3.5in]{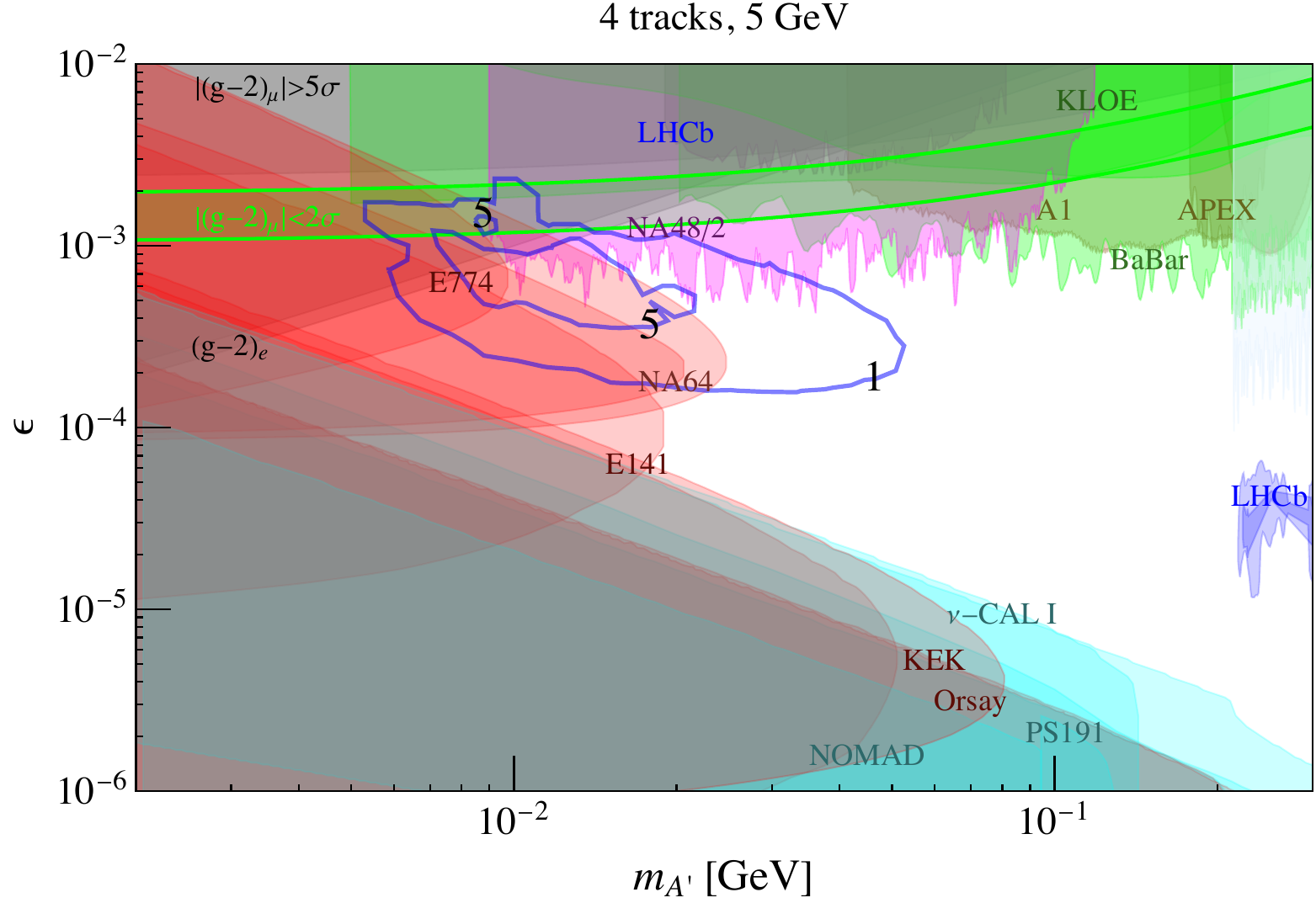}
  \caption{Contours of the total number of signal events under different search strategies.
  }
  \label{fig:contours}
\end{figure*}
\section{Contours of number of events for different searches}
\label{sec:contours}

Here we show the total expected number of signal events for different searches in~\figref{contours}.

\bibliography{dark_photons_at_MUonE}

\end{document}